\documentclass[sigconf,authorversion,nonacm,10pt]{acmart}

\AtBeginDocument{%
  }

\setcopyright{acmlicensed}
\copyrightyear{2018}
\acmYear{2018}
\acmDOI{XXXXXXX.XXXXXXX}

\acmConference[Conference acronym 'XX]{Make sure to enter the correct
  conference title from your rights confirmation emai}{June 03--05,
  2018}{Woodstock, NY}
\acmISBN{978-1-4503-XXXX-X/18/06}

\usepackage{graphicx}  
\usepackage{caption}
\usepackage{multirow}
\begin{document}

\title{DIETS: Diabetic Insulin Management System in Everyday Life}



\author{Zeng, Hanyu}
\email{Haz207@pitt.edu}
\authornotemark[1]
\affiliation{%
  \institution{University of Pittsburgh}
  \city{Pittsburgh}
  \state{Pennsylvania}
  \country{USA}
  \postcode{15213}
}
\author{Ji, Hui}
\email{huj16@pitt.edu}
\authornotemark[1]
\affiliation{%
  \institution{University of Pittsburgh}
  \city{Pittsburgh}
  \state{Pennsylvania}
  \country{USA}
  \postcode{15213}
}
\author{Zhou, Pengfei}
\email{pengfeizhou@pitt.edu}
\authornotemark[1]
\affiliation{%
  \institution{University of Pittsburgh}
  \city{Pittsburgh}
  \state{Pennsylvania}
  \country{USA}
  \postcode{15213}
}

\renewcommand{\shortauthors}{Anonymous}

\begin{abstract}
People with diabetes need insulin delivery to effectively manage their blood glucose levels, especially after meals, because their bodies either do not produce enough insulin or cannot fully utilize it. Accurate insulin delivery starts with estimating the nutrients in meals and is followed by developing a detailed, personalized insulin injection strategy. These tasks are particularly challenging in daily life, especially without professional guidance. Existing solutions usually assume the prior knowledge of nutrients in meals and primarily rely on feedback from professional clinicians or simulators to develop Reinforcement Learning-based models for insulin management, leading to extensive consumption of medical resources and difficulties in adapting the models to new patients due to individual differences.
In this paper, we propose DIETS, a novel diabetic insulin management framework built on the transformer architecture, to help people with diabetes effectively manage insulin delivery in everyday life. Specifically, DIETS tailors a Large Language Model (LLM) to estimate the nutrients in meals and employs a titration model to generate recommended insulin injection strategies, which are further validated by a glucose prediction model to prevent potential risks of hyperglycemia or hypoglycemia.
DIETS has been extensively evaluated on three public datasets, and the results show it achieves superior performance in providing effective insulin delivery recommendation to control blood glucose levels. 

\end{abstract}

\maketitle

\section{Introduction}
\label{sect:introduction}

\begin{figure}
    \centering
    \includegraphics[width=0.46\textwidth]{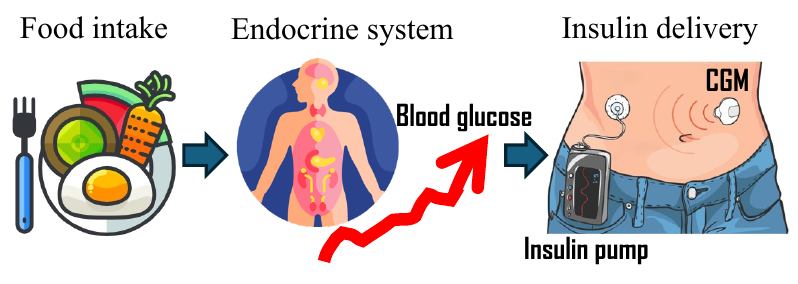}
    \caption{Insulin delivery for people with diabetes.}
    \label{Diabetes}
\end{figure}

Diabetes is emerging as one of the most significant global epidemics, affecting over 10\% of the adult population, with this number on the rise \cite{zimmet2017diabetes}. The high blood glucose level is primarily due to autoimmune destruction of insulin-producing $\beta$ cells or the development of insulin resistance. This necessitates sophisticated insulin therapy regimens that mimic the body's natural insulin release, involving continuous basal secretion complemented by larger bolus doses during meals to regulate glucose levels \cite{jaloli2024basal, cescon2020using}.
As illustrated in Figure \ref{Diabetes}, an appropriate insulin delivery strategy is primarily determined by the patient's dietary intake, individual insulin sensitivity, and the regulation of the endocrine system \cite{american20216, rodbard2011glycemic}.
For example, high-carbohydrate meals demand rapid insulin administration, while meals rich in proteins and fats require slower, more prolonged insulin infusions due to delayed glucose metabolism \cite{evert2020factors}. 
Effective delivery strategy determination requires a deep understanding and analysis of both clinical and nutritional sciences, highlighting the indispensable role of professional expertise. 

Accurate insulin delivery begins with estimating the nutrients in a patient's meal and is followed by developing a detailed titration strategy based on the patient's endocrine system.
These two tasks, however, are non-trivial in practice. First, analyzing the nutrients in meals necessitates a deep understanding of nutritional science and access to a comprehensive nutrition knowledge base to accurately quantify the various nutrients in the consumed food.
Second, understanding how these nutrients affect blood glucose levels and the role of insulin in the endocrine system in regulating these levels are crucial. However, the complexity of the human endocrine system make it difficult to precisely quantify the relationship between insulin and glucose, and variations in physiological conditions over time and across individuals further complicate the task for non-professionals to determine the appropriate insulin dosage.
For most patients, continuous access to professional medical guidance for insulin administration is impractical, making precise insulin titration in everyday life a critical and daunting task.

Existing approaches struggle to effectively address these two tasks in everyday situations. 
For the first task, unsupervised learning models are used to analyze the textual nutritional data \cite{ispirova2020p,ispirova2024msgen}, but they fail to handle unstructured patient descriptions. Thus most research works either assume having prior knowledge about the nutrients of the patient's meals \cite{brown2018first, silva2022real} or require the patient to adhere to a predefined set of food intake \cite{emerson2023offline,jaloli2024basal}. However, in everyday life scenarios, such assumptions and requirements are impractical, leaving patients to rely largely on their own rough estimates, which is usually inaccurate due to the variability in daily meals and the absence of professional guidance.
For the second task, the widespread adoption of Continuous Glucose Monitoring (CGM) systems \cite{CGM, ahmadi2009wireless} has catalyzed the development of data-driven approaches, which generally fall into two categories: traditional control theory based approaches and AI-driven methods. Traditional approaches, such as Model Predictive Control (MPC) and Proportional-Integral-Derivative (PID), use real time CGM data to adjust insulin injections \cite{el2009review}. However, the significant inherent delay (e.g., 30 minutes \cite{galloway1981factors}) in insulin's impact on blood glucose levels makes it hard to control glucose levels using control theory based approaches, often resulting in hyperglycemia. 
Consequently, more sophisticated AI methods, particularly Reinforcement Learning (RL), have been introduced \cite{wang2023optimized,yu2021reinforcement,yau2023reinforcement}, where professional clinicians assess and provide feedback to guide the model's decisions, allowing the model to develop professional decision-making skills. However, these approaches consume extensive medical resources and struggle with adapting the model to new patients due to individual differences among patients \cite{wexler2022patient}. 

\begin{figure}
    \centering
    \includegraphics[width=0.45\textwidth]{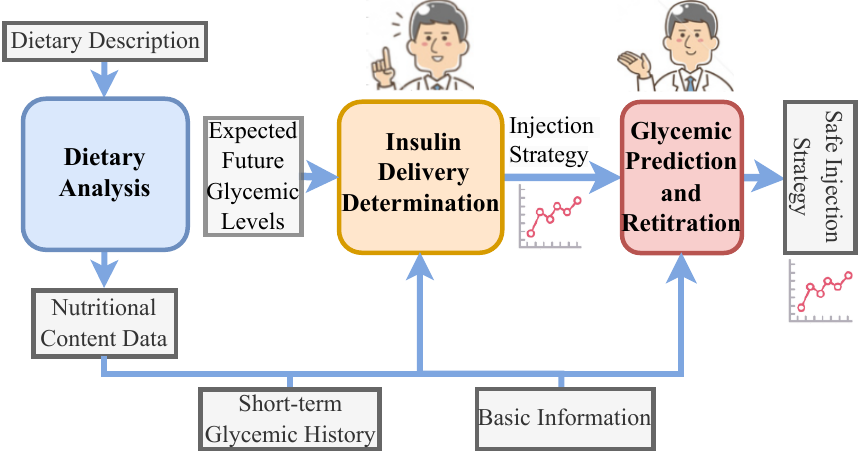}
    \caption{The system workflow of DIETS.}
    \label{system}
\end{figure}

This paper is motivated by an essential question: \textit{Is it possible to achieve accurate blood glucose control for people with diabetes in everyday life without the need of expert supervision?}
To achieve this goal, in this paper, we propose a novel \textbf{D}iabetic \textbf{I}nsulin manag\textbf{E}men\textbf{T} \textbf{S}ystem designed to provide precise and safe insulin titration for people with diabetes in everyday life. DIETS consists of three main modules as shown in Figure \ref{system}: dietary analysis, insulin delivery strategy determination, and glucose prediction with re-titration capabilities. 
Patients input the description of their past and projected dietary intake, based on which a tailored Large Language Model (LLM) estimates the nutrients, including calories, carbohydrates, proteins, and fats. 
This nutritional information, along with the expected glucose levels and the other essential information such as the patient's prior insulin injections and basic personal information, are fed into the insulin delivery strategy determination module to generate recommended dosages for a forthcoming period. 
These recommendations are then evaluated by the glucose prediction model to ensure the proposed insulin strategies are safe. If potential risks of hyperglycemia or hypoglycemia are detected, a tailored LLM re-evaluates the situation to generate revised insulin strategies, which undergo another round of risk assessment in the glucose prediction model until a sufficiently safe insulin dosing plan is established.

DIETS has the following advantages.
First, DIETS employs a two-stage training process, where a foundation system is developed using data records in public datasets in the first stage and the model is fine-tuned with minimal inputs from the individual patient in the second stage. By doing so, DIETS is able to be easily adapted to different individuals while maintaining high performance.
Second, unlike traditional AI models, LLMs possess deeper and broader knowledge bases. We adapt the LLM with in-context learning \cite{Dong2024InContextLearning} to perform accurate dietary nutrition analysis and significantly reduce patients' reliance on professionals.
Third, unlike previous works that rely on expert feedback to make decisions, the insulin delivery strategy determination module learns solely from the patient's glycemic data on how insulin and glucose levels influence each other. We train the model using all data in the records by including the records where the glucose levels are out of the safe range. By doing so, the transformer-based model is able to better capture the relationship and interactions between insulin and blood glucose.
Forth, the glucose prediction component posses as a safety mechanism that predicts whether the recommended dosing strategy will lead to unsafe glucose fluctuations and re-titration if there is a risk of hazardous events. 

We evaluate DIETS' performance on three public clinical datasets \cite{zhao2023chinese, marling2020ohiot1dm}. 
As DIETS is the first comprehensive framework that spans the entire process of insulin management in everyday life, including nutrient estimation, and insulin delivery determination and glucose prediction, we compare the performance of individual modules used in DIETS with related existing solutions, respectively.
Specifically, first, we experiment with different LLMs for the dietary analysis. All tailored LLMs show promising results, with GPT-4o \cite{gpt4o2024} and Mistral-22B \cite{jiang2024mixtral} achieving marginally superior performance, which is much better than existing methods.
Second, we compare the quality of the insulin delivery strategies determined by DIETS with those from the latest state-of-the-art (SOTA) solution. As all SOTA solutions assume prior knowledge about the nutrients of the patient’s meals, we use the nutrient estimation from our dietary analysis module as inputs for all models. The experiments demonstrate that DIETS produces the most effective insulin delivery strategies, achieving at least 50\% reduction in the mean absolute error.
Third, we compare the performance of the blood glucose prediction model with SOTA models in the literature, and DIETS also achieves the best performance, reducing the mean error by more than 25\%.

Our main contributions are summarized as follows:
\begin{itemize}
\item We introduce the first comprehensive insulin management framework, DIETS, for people with diabetes in everyday life without the need of expert supervision. DIETS enables patients to receive effective insulin delivery recommendations, which are verified by a protection mechanism.
\item DIETS innovates across different modules within the framework. By utilizing a tailored LLM for dietary analysis, patients can obtain accurate nutrient estimation from simple descriptions of their meals, which significantly reduces the need for professional intervention. With minimal personal data to fine-tune the system, our strategy model is capable of determining safe and precise injection recommendations. The proposed glucose prediction model incorporates a deeper analysis of individual patient information to provide customized glucose forecasts, and re-titrate when potential hazards are predicted.
\item We conduct extensive experiments to evaluate DIETS on three public datasets. The results suggest that DIETS outperforms SOTA models in all modules, demonstrating its superior performance in these critical aspects of diabetes management.
\end{itemize}

\section{Background and Motivation}
\label{sect:background}

\subsection{Diabetes}
\label{subsect:diabetes}
Diabetes has become a major global epidemic, affecting 10.5\% of adults aged 20-79 affected worldwide, totaling 537 million individuals \cite{zimmet2017diabetes}. Projections suggest that these numbers will increase to 643 million by 2030 and 783 million by 2045 \cite{kumar2024prevalence}. Diabetes is not only prevalent but also a serious health threat, causing an estimated 6.7 million deaths in 2021 alone.
The condition is categorized into two main types: Type 1 and Type 2. Type 1 diabetes (T1D) is an autoimmune metabolic disorder characterized by the destruction of $\beta$ cells, leading to insufficient insulin production and resultant hyperglycemia \cite{katsarou2017type}. In contrast, Type 2 diabetes (T2D) initially involves normal insulin production, but patients develop resistance to it over time, causing glucose levels to rise dangerously high. Such elevated glucose levels can be toxic, potentially leading to serious complications including diabetic retinopathy, neuropathy, cardiovascular diseases, and limb loss \cite{dhatariya2020diabetic}.
Given these severe complications, it is crucial to manage diabetes carefully. One aspect of such management is insulin dosing. Improper insulin dosing can lead to blood glucose instability \cite{chin2021effects} and hypoglycemia, which can cause some other severe complications even immediate death  \cite{frier2011hypoglycemia,kalra2013hypoglycemia}. These critical risks underscore the importance of meticulous glycemic management for people with diabetes.

\subsection{Insulin Delivery}
\label{subsect:insulin_delivery}
Insulin delivery is typically administered through basal-bolus therapy. Basal insulin, which is usually stable, balances the blood sugar fluctuations caused by the daily metabolic (during fasting) process. In contrast, blood glucose fluctuations caused by ingested nutrients are mainly adjusted by bolus insulin supplementation \cite{king2005basal}.
Carbohydrates, proteins, and fats are key nutrients that impact blood glucose levels. Carbohydrates are the primary drivers of blood glucose fluctuations \cite{deeb2017accurate}, as they are quickly metabolized into glucose. Proteins and fats, however, have a more gradual effect. Specifically, proteins can induce blood glucose changes 2-5 hours post-meal by affecting hormone secretion \cite{lowe2008flexible, dafne2002training, diabetes1993effect}, while fats slow the metabolic breakdown of carbohydrates, moderating blood glucose rises and delaying peaks until 3-5 hours after eating \cite{paterson2015role}.
Standard short-acting insulin, which peaks between 80-120 minutes after injection, may not adequately address the extended blood glucose elevations caused by high-protein and high-fat meals \cite{wong2021ultra}. Considering these nuances, insulin pumps with square or dual wave functions allow for prolonged insulin delivery, aligning insulin activity with delayed glucose rises from different types of meals \cite{heinemann2009insulin}, thereby helping to prevent postprandial hyperglycemia.
This understanding motivates us to customize insulin delivery strategies based on meal composition.
Instead of a one-time injection dosage guidance, we may need to provide a long-term injection strategy recommendation, which can give patients adequate psychological preconception and make the injection more smooth.

\subsection{Commercial Insulin Delivery System}
\label{subsect:commercial_insulin_delivery}
The commercial automated insulin delivery (AID) system (Figure \ref{Diabetes}), such as the Omnipod 5 \cite{Omnipod5}, typically consists of an insulin pump, a CGM, and a control theory based algorithm \cite{nwokolo2023artificial} running on smartphones. The CGM measures the glucose concentration every 5-10 minutes. The smartphone, using a control theory-based algorithm, calculates the required insulin dosage and adjusts the insulin pump delivery through a subcutaneous tube accordingly. Due to the lack of nutrient intake estimation and the significant inherent delay in insulin's impact on blood glucose levels, which is approximately 30 minutes \cite{galloway1981factors}, AID is far from perfect even though it made a positive impact on diabetes management.

\subsection{Motivation}
\label{subsect:motivation}

To develop a robust blood glucose management system to help people with diabetes in everyday life, two key tasks must be addressed: accurately estimating the nutrients in meals and determining effective and safe insulin dosages tailored to the individual patient's condition.

First, accurate nutrient estimation is typically a task for professionals, given its requirement for a vast nutritional knowledge base which poses a challenge for ordinary patients. Large language models (LLMs) represent a significant advancement in artificial intelligence \cite{achiam2023gpt,team2023gemini,touvron2023llama}. They are trained on trillions of language tokens, amassing extensive knowledge bases that enable them to understand and process natural language effectively. With billion-level parameters, these models possess profound analytical and inferential capabilities in natural language tasks. We may tailor LLMs with simple in-context learning methods  \cite{Dong2024InContextLearning} to adapt them to accurately estimate nutrients from the patient's unstructured dietary descriptions.

Second, recent studies have applied Reinforcement learning (RL) \cite{kaelbling1996reinforcement} based models for blood glucose control \cite{wang2023optimized, jaloli2024basal,emerson2023offline} utilizing iterative feedback from professionals to refine decision-making capabilities. These RL models are typically trained within simulators \cite{jaloli2024basal,emerson2023offline} to circumvent the ethical and practical challenges associated with conducting direct experiments within the human body \cite{griesdale2009intensive}. However, these simulators often fall short in capturing the full complexity of human physiological responses, which limits the realism and applicability of the feedback used for training. A data-driven method introduced in \cite{wang2023optimized} involves real-world data records from hospitalized patients, where decisions made by RL models are evaluated by experienced clinicians. While this method enables the immediate application of insights generated by RL, it heavily depends on the availability of experienced medical personnel for supervision. This dependence creates a bottleneck, limiting the method's scalability and generalizability, hindering its widespread adoption in everyday life. Instead, we focus on developing AI models to understand the relationship between insulin injections and blood glucose fluctuations, based on which the model can mimic human physiological responses to enable more accurate insulin administration for a target glucose level.

Additionally, severe fluctuations such as hyperglycemia and hypoglycemia have serious health implications, necessitating an additional glucose prediction component as a safety measure to prevent potential risks. However, current glucose prediction solutions \cite{li2019convolutional, yang2023short} overlook the personalized information of individuals, resulting in unstable performance in predicting blood glucose levels. To achieve accurate prediction, we need to incorporate patient-specific data to tailor the model for different individuals.


\section{DIETS Design}
\label{sect:DIETS_design}
\begin{figure*}
    \centering
    \includegraphics[width=0.9\textwidth]{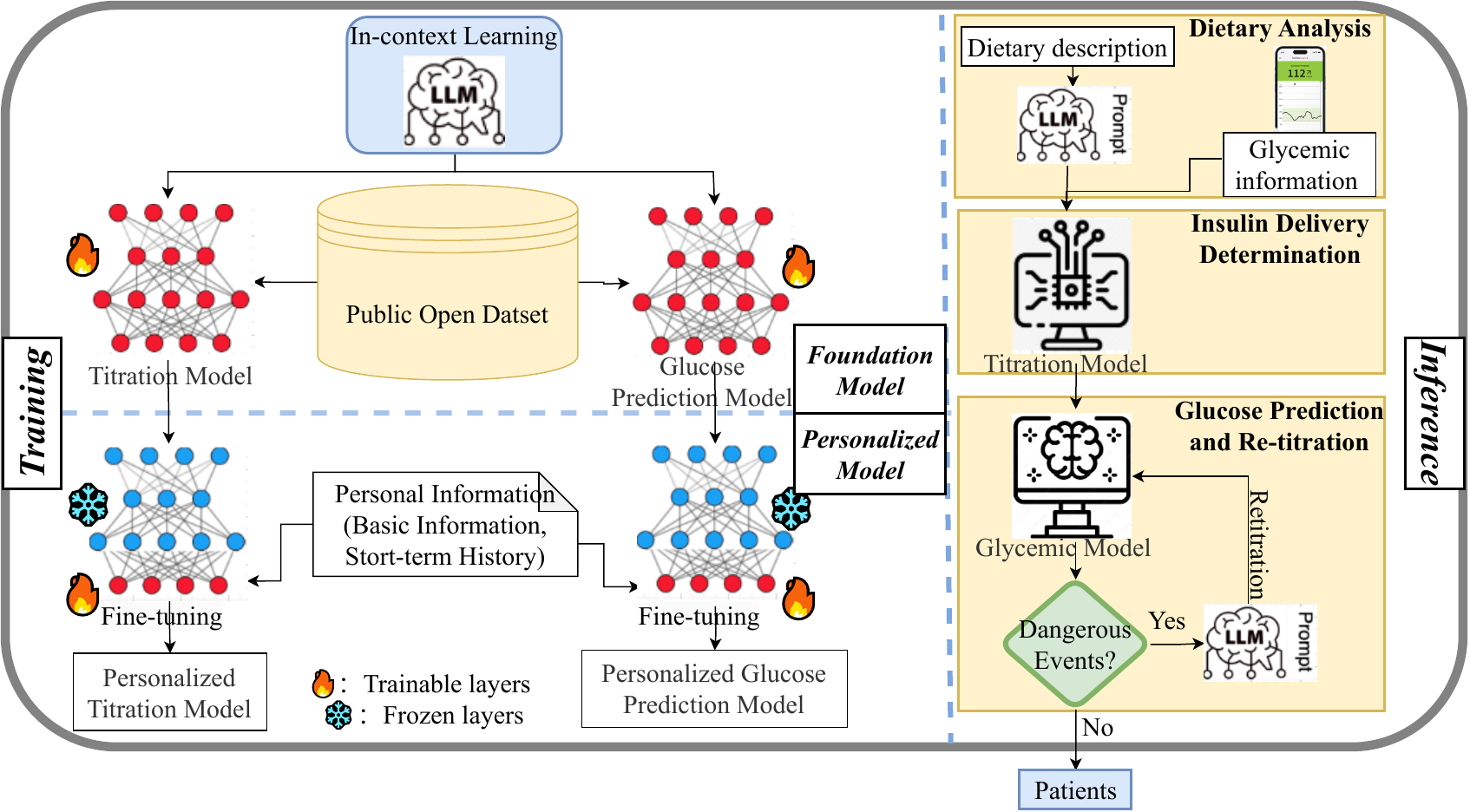}
    \caption{The overview structure of DIETS for training and inference.}
    \label{infer_train}
\end{figure*}
Inspired by the mentioned motivation, in this paper, we propose a novel diabetic insulin management framework, \textbf{DIETS}, to help diabetes patients effectively manage their blood glucose levels with appropriate insulin delivery. In this section, we present the detailed design of DIETS, starting with an overview of the framework and then delving into the design of each component.

\subsection{Overview}
\label{subsect:DIETS_overiew}

As shown in Figure \ref{infer_train}, the framework includes three components, i.e., dietary analysis module, 
insulin delivery determination module, and glucose prediction and re-titration module.
DIETS focuses on the bolus insulin injections because basal insulin dosages are relatively stable, and easy to manage based on long-term physiological conditions. 
The dietary analysis module takes the input of descriptions of the consumed meal provided by the patient. Using a tailored LLM, it derives the nutritional information of the meal such as calorie content, carbohydrate, protein, and fat. This nutritional data is then utilized by the insulin delivery determination module to develop insulin injection strategies. Instead of one-dose injection, this module determines a multi-dose insulin injection strategy which spans the next 2 hours.
Following the determination, the glucose prediction and re-titration module is employed to forecast future glucose levels in conjunction with the administered insulin dosage, preventing the occurrence of serious events such as hyperglycemia and hypoglycemia. If a potential risk is identified, the system will generate a new titration recommendation until the proposed injection regimen is deemed safe.

\subsection{Data Normalization and Segmentation}\label{dp}
Before conducting analysis, a series of data pre-processing steps are necessary. To keep it consistent with the CGM, most data are sampled every 15 minutes, including dietary intake, blood glucose levels, insulin injections, and anti-diabetic drug intake. All types of data are normalized to uniform units: insulin doses in insulin units ($IU$)\footnote{1 insulin unit ($IU$) = 0.01 $mL$.}, blood glucose in milligrams per deciliter ($mg/dl$), nutrient and anti-diabetic drug intake in grams ($g$), and energy in calories ($cal$). Following normalization, all time-series data are concatenated into a single data-trace. We employ a sliding window of width $m$ to segment the data trace into discrete clips $B_i$ comprising $B_1, B_2, ..., B_m$. Any missing values in these segments were filled with zeros. Each data clip contained data from $m$ time points, where the initial $n$ points were categorized as 'previous' data and the remaining $m-n$ points as 'future' data. Future insulin injection data and glucose levels are masked during training of the titration and glucose prediction model and used as labels.



\subsection{Dietary Analysis}
\begin{figure}
    \centering
    \includegraphics[width=0.5\textwidth]{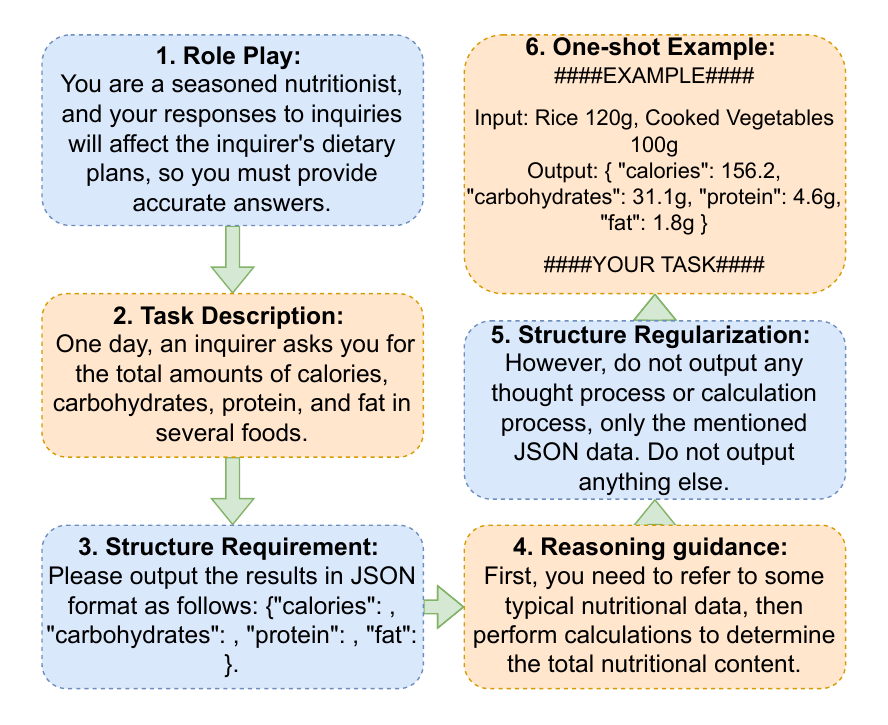}
    \caption{The prompt instance for dietary analysis.}
    \label{prompt}
\end{figure}

The dietary analysis module provides nutritional content estimation for the following insulin delivery determination module and glucose prediction and re-titration module.
The dietary analysis module processes patient-provided dietary descriptions through a Large Language Model (LLM). We tailored the LLM using in-context learning with a specifically designed prompt to learn to analyze the nutrient content of consumed diets for subsequent insulin titration. The LLM's robust few-shot learning capabilities allow it to effectively parse and interpret unstructured dietary descriptions, which traditional NLP models may find challenging due to variable text lengths and formats. An appropriately crafted prompt enhances the model’s accuracy by assigning the LLM the role of nutritionist and detailing the analysis task with a focus on outputting structured data for easy integration into subsequent processes. 
Figure \ref{prompt} illustrates a prompt instance when tailoring the LLM into the role of a nutritionist. We include the following components in the prompt to fine-tune the LLM: role play instruction, task description, structure requirement on the output format, reasoning guidance on how  to estimate the nutrients, structure regularization, and a one-shot example.

\subsection{Insulin Delivery Determination}
The Insulin Strategy Determination module aims to determine a optimal insulin dosing strategy for the patient to achieve the anticipated changes in blood glucose levels in the next 2 hours. In DIETS, we leverage the following medical records: historical blood glucose fluctuations and insulin injection data, dietary nutrient information from dietary analysis, and etc. The key is to learn the relationship and interaction between insulin and blood glucose levels.

\begin{figure}
    \centering
    \includegraphics[width=0.5\textwidth]{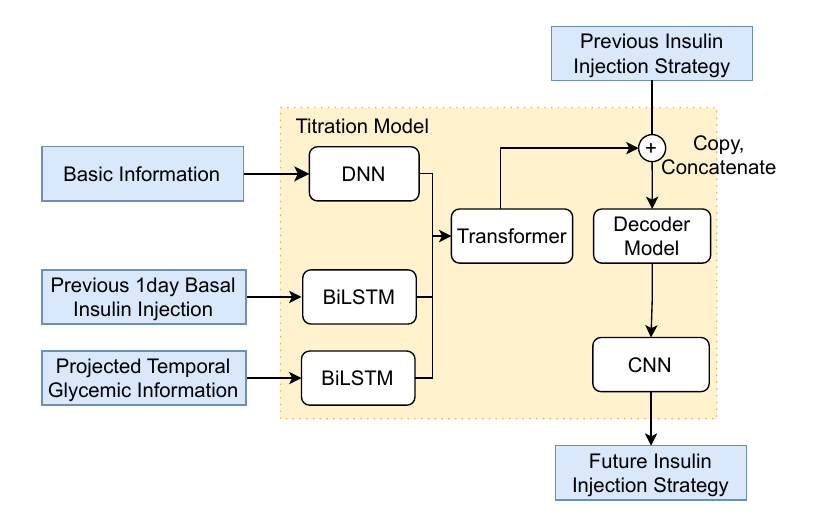}
    \caption{The structure of the titration model.}
    \label{titration_m}
\end{figure}

As depicted in Figure \ref{titration_m}, the insulin strategy determination model inputs both past and projected glycemic-related information into a BiLSTM to extract temporal features. This includes nutritional intake estimations, anticipated blood glucose levels, anti-diabetic drug intake, and the projected basal insulin injection. Personal basic information is processed through a Deep Neural Network (DNN) to extract latent features. Data on basal insulin injections from the previous day, which partly represent the patient's recent physiological condition, are also considered. The outputs from these three processes are then fed into a transformer-based model. The results are structured to align with the data format of past insulin injection strategies, concatenated, and then input into a Generative Pre-trained Transformer (GPT)-like decoder \cite{yenduri2023generativepretrainedtransformercomprehensive}. Subsequently, a CNN processes these inputs sequentially to generate output results. 
The formula for this module is as follows.
\begin{align*}
X = \text{Transformer}(
    \text{concat}(
    \text{DNN}(B_I), 
    \text{BiLSTM}(B_a^{(24h)}), &\\
    \text{BiLSTM}(T_{[1,...,m]})
    )
) \\
B_{O[n+1]} = \text{CNN}(
\text{Decoder}(
    \text{concat}(
    \text{X.copy}(n), 
    B_{O[1,...,n]}))),
\end{align*}
where $X$ represents the outputs of the transformer, $B_I$ is the basic information of patients, and $B_a^{(24h)}$ is the basal injection of the last 24 hours, and $T_{[1,...,m]}$ represents the temporal glycemic information involving the glucose levels, nutrient intake, and anti-glycemic drug intake. In the second formula, $B_{O}$ is the bolus injection dosages, and $B_{O[1,...,n]}$ represents the historical bolus injection dosages, and $B_{O[n+1]}$ is the bolus injection recommended for the future.

During training, insulin dosages from the last two hours in the cropped data segments are used as labels, with all other relevant patient data serving as inputs. This training process enables the titration model to learn the effects of various insulin injection strategies on blood glucose control across endocrine systems of different patients. 
It is worth mentioning that in the training process, we also include the records where the glucose levels are out of the safe range, e.g., higher than 180 $mg/dl$ or lower than 70 $mg/dl$. By doing so, the model is able to better capture the relationship and interactions between insulin and blood glucose.
In the inference phase, patients can set their desired glucose levels for the next 2 hours, and the model uses its learned knowledge to propose an appropriate insulin injection strategy for the next 2 hours with 8 dosages (15 minutes interval).

\subsection{Glucose Prediction and Re-titration}
This module serves as a safe guardian to prevent potential risks like hypoglycemia and hyperglycemia by predicting and re-titrating. 
Hypoglycemia generally refers to a blood glucose level that falls below the normal range of 65-70 $mg/dl$ (3.6–3.9 $mmol/l$) \cite{cryer2003hypoglycemia}. Hypoglycemia is extremely dangerous and can lead to serious conditions such as ataxia, mental confusion, speech impairments, seizures, coma, and in the most severe cases, death \cite{agrawal2022impact}. 
Hyperglycemia is defined as a fasting blood glucose level exceeding 125 $mg/dl$, or a postprandial (2 hours after meal) blood glucose level exceeding 180 $mg/dl$. If left untreated, it can lead to many serious and life-threatening complications, including damage to the eyes, kidneys, nerves, heart, and peripheral vascular system \cite{mouri2023hyperglycemia}.

\subsubsection{Glucose Prediction}

\begin{figure}
    \centering
    \includegraphics[width=0.5\textwidth]{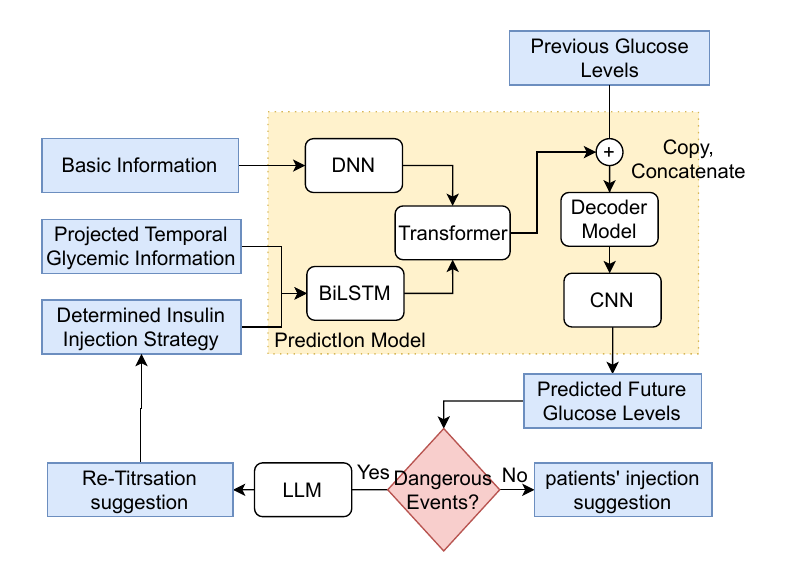}
    \caption{The structure of the hyperglycemia and hypoglycemia events detection and re-titration system.}
    \label{gp}
\end{figure}

The glucose prediction model serves to predict blood glucose concentrations. As illustrated in Figure \ref{gp}, it shares a structural resemblance with the titration model shown in Figure \ref{titration_m}; both are GPT-like decoder models but differ in the type of input information they process. In the prediction model, projected time-series data, including anticipated nutritional intake, anti-diabetic drug consumption, and the projected basal insulin injections in the time under consideration, are input into a BiLSTM layer to extract temporal features. These features are combined with patient characteristics extracted by a DNN and fed into a transformer layer. After alignment with past blood glucose data through replication and concatenation, the inputs are processed by a GPT-like decoder model. Finally, a CNN layer sequentially generates predictions for future glucose levels. The formula for the glucose prediction model is as follows.
\begin{align*}
\text{x} = \text{Transformer}(
    \text{concat}(
    \text{DNN}(B_I),
    \text{BiLSTM}(t_{[1,...,m]}))
    ) 
),
\end{align*}
\begin{align*}
G_{[n+1]} = \text{CNN}(
\text{Decoder}(
    \text{concat}(
    \text{x.copy}(n), 
    G_{[1,...,n]})
    ) 
),
\end{align*}
where $x$ represents the outputs of the transformer, $B_I$ is the basic information of patients, and $t_{[1,...,m]}$ represents the glycemic information involving the insulin injection dosages, nutrient intake, and anti-glycemic intake. In the second formula, $G$ is the glucose level, and $G_{[1,...,n]}$ represents the glucose levels for previous time points, and $G_{[n+1]}$ is the glucose level for the next time point.

During training, similar to the titration model, we use glucose changes over the last two hours to train this model. The actual insulin injection strategies from the preceding two hours, along with all other relevant information within the data clips, are used to predict blood glucose changes.
During inference, this model forecasts the potential glucose changes that the proposed insulin delivery strategies might induce, thereby assessing whether the strategies are sufficiently safe and rational for implementation.
The criterion for identifying potential issues involves checking whether the predicted glucose concentrations are below 70 $mg/dl$ or exceeding 180 $mg/dl$ during the forecast period. 
If a potential risk is identified, the predicted insulin dosages are referred back to do re-titration until all risks are mitigated. Otherwise, this safe insulin injection strategy is provided to the patient.

\subsubsection{Re-Titration}

A LLM is employed to facilitate the re-titration process through carefully designed prompts. In the prompt, the LLM assumes the role of a diabetes specialist. 
The model is instructed that increased dosages may be necessary to manage hyperglycemia and reduced dosages for hypoglycemia. The prompt details the current glucose levels, recent dietary intake, anticipated insulin injections, and prediction of affected blood glucose levels. 
Once the LLM's re-titration suggestions are obtained, these new dosing recommendations are input into the glucose prediction model for a subsequent round of event detection, assessing potential hyperglycemia and hypoglycemia. 
\subsection{DIETS-LSTM}
In DIETS, the decoder layers of the diet titration and glucose prediction models predominantly consist of stacked transformers, favored for their proficiency in capturing correlations between diverse types of data. This ability makes them particularly effective at learning from multimodal information when patients provide various personal details. However, when such comprehensive personal information is unavailable, transformers may not perform optimally and could introduce unnecessary complexity. We opt for LSTM layers to form the decoder layer in such scenarios to address this. This adjustment reduces model complexity and enhances accuracy. We refer to this specific configuration as DIETS-LSTM.

\subsection{Two-stage Training}

To train patient-specific models without relying on professional guidance, the titration model and glucose prediction model are trained separately, and both employ a two-phase training, as shown in Figure \ref{infer_train}.
In the first phase, we train a foundation model using public open datasets, allowing it to learn generalizable knowledge about human physiological and metabolic hormone responses. In the second phase, this foundation model is fine-tuned using a small amount of information about the patient, including the patient's basic information and few days (e.g., 3 days) historical glycemic data. 
The fine-tuning process specializes the generalizable knowledge learned to fit the specific individual’s context. During this phase, only the final dense layer of the model is fine-tuned, and other layers are frozen to preserve the knowledge acquired from the extensive datasets. 

Two-stage training ensures comprehensive learning, deriving insights from basic patient data and historical medical records in the datasets. Consequently, the model not only learns complete information necessary for precise insulin titration but also utilizes a synthesis of personal health data and temporal dynamics to enhance diabetes management, culminating in a GPT-like decoder that sequentially generates insulin dosages for the upcoming two hours based on previous dosages and patient-specific information.

\section{Experiment}
\subsection{Datasets and Implementation Details}
\label{subsect:experiment_details}

\textbf{Datasets.} Our experiments are conducted on three public datasets. We use the OhioT1DM dataset \cite{marling2020ohiot1dm} and the ShanghaiT1DM and ShanghaiT2DM datasets \cite{zhao2023chinese} for evaluating the titration model and glucose prediction model in DIETS. 
For simplicity, we refer to the ShanghaiT1DM and ShanghaiT2DM datasets as ShanghaiDM dataset in the following.
The dietary analysis module is evaluated with the BOOHEE dataset\footnote{https://www.boohee.com/}. The \textbf{ShanghaiDM dataset}, introduced in 2023, encompasses data from 12 Type 1 and 100 Type 2 diabetic patients, including basic demographics like height, weight, age, sex, BMI, duration of illness, and lifestyle factors such as smoking and drinking habits. It also involves detailed medical records, consisting of parameters such as Fasting Plasma Glucose, Fasting C-peptide, 2-hour Postprandial C-peptide, Fasting Insulin, HbA1c, Glycated Albumin, repeated measurements of Total Cholesterol, HDL, LDL, Creatinine, eGFR, Uric Acid, BUN, and a Hypoglycemia indicator. Each patient's data features a 3$\sim$14-day record of blood glucose fluctuations, bolus, and basal insulin injections, textual dietary intake descriptions, and anti-glycemic drug intake, recorded every 15 minutes. 11 Patients of the T1 dataset use Novolin insulin and the rest one Type1 patient uses both Humulin and Gansulin insulin. The \textbf{OhioT1DM dataset}, released in 2018, details 8 weeks of data from 6 Type 1 diabetic patients, documenting blood glucose levels, insulin injections, estimated carbohydrate intake per meal, along with records of physical activity, work, sleep, and additional metrics such as stress, heart rate, temperature, and step count, captured every 5 minutes. All patients in the OhioT1dm dataset use Novalog insulin. The \textbf{BOOHEE dataset}, which includes nutritional content for over 4000 foods, recipes, and reviews, is utilized in our experiments primarily to assess the model's performance with home-cooking data, offering a comprehensive test of its dietary analysis capabilities in realistic settings.

\textbf{Data pre-processing.} For the ShanghaiDM dataset, we perform data cleaning such as converting textual descriptions of insulin injections into numerical values. The dataset details insulin injections administered via both subcutaneous and intravenous routes. Given that subcutaneous injections typically delay insulin absorption by about half an hour \cite{galloway1981factors}, we adjusted these data entries to align the timing effects of both administration methods. 

\textbf{Train Setup.} All the experiments are conducted on 1 NVIDIA A100 GPU, utilizing a batch of 32. The AdamW algorithm governs the optimization process. This configuration is sustained throughout 200 epochs with 40 epochs' patience early stopping. The learning rate is set at 0.005. The distribution of data across the training, validation, and test sets is configured to be 70\%, 15\%, and 15\%, respectively.

\textbf{Models in comparison:} We compare the performance of DIETS with the following state-of-the-art models. 
\textbf{P-Nut} \cite{ispirova2020p} is a machine learning pipeline for predicting macronutrient values of foods using unsupervised learning for clustering, followed by supervised learning for regression.
\textbf{RL-DITR} \cite{wang2023optimized} is a reinforcement learning (RL) based model trained based on the professionals' instructions and case-specific feedback. It is the latest data-driven instead of simulator-trained RL model in the literature. 
\textbf{PGBTAM} \cite{yang2023short} is a short-term glucose prediction model based on the temporal multi-head attention mechanism. 
\textbf{CRNN} \cite{li2019convolutional} uses Convolutional Recurrent Neural Networks for glucose prediction. 
\textbf{P-LSTM} \cite{prendin2023importance} is a LSTM-based model for blood glucose prediction.

\subsection{Evaluation Metrics}
We evaluate model performance using Mean Absolute Error (MAE) and variance. The performance of dietary analysis, titration model, and glucose prediction model are evaluated by MAE, and the variance is used to evaluate the stability of the tailored LLM on dietary analysis task. $\text{MAE} = \frac{1}{n} \sum_{i=1}^{n} \left| y_i - \hat{y}_i \right|
$, where $n$ is the number of data, and $y_i$ is the true value for the $i$th data, and the $\hat{y}_i$ is the calculated value of the evaluated model. Variance $\sigma^2 = \frac{1}{n} \sum_{i=1}^{n} \left( x_i - \mu \right)^2$, where $n$ is the time of experiment, and $x_i$ is the calculated value of the $i$th experiment, and $\mu$ represents the mean value of $x$ in all the experiments.

\section{Evaluation}
This section presents the experiment results. DIETS is the only comprehensive framework that spans the entire process of insulin management. Thus we compare the performance of the modules used in DIETS with related existing solutions.

\begin{figure}
    \centering
    \includegraphics[width=0.48\textwidth]{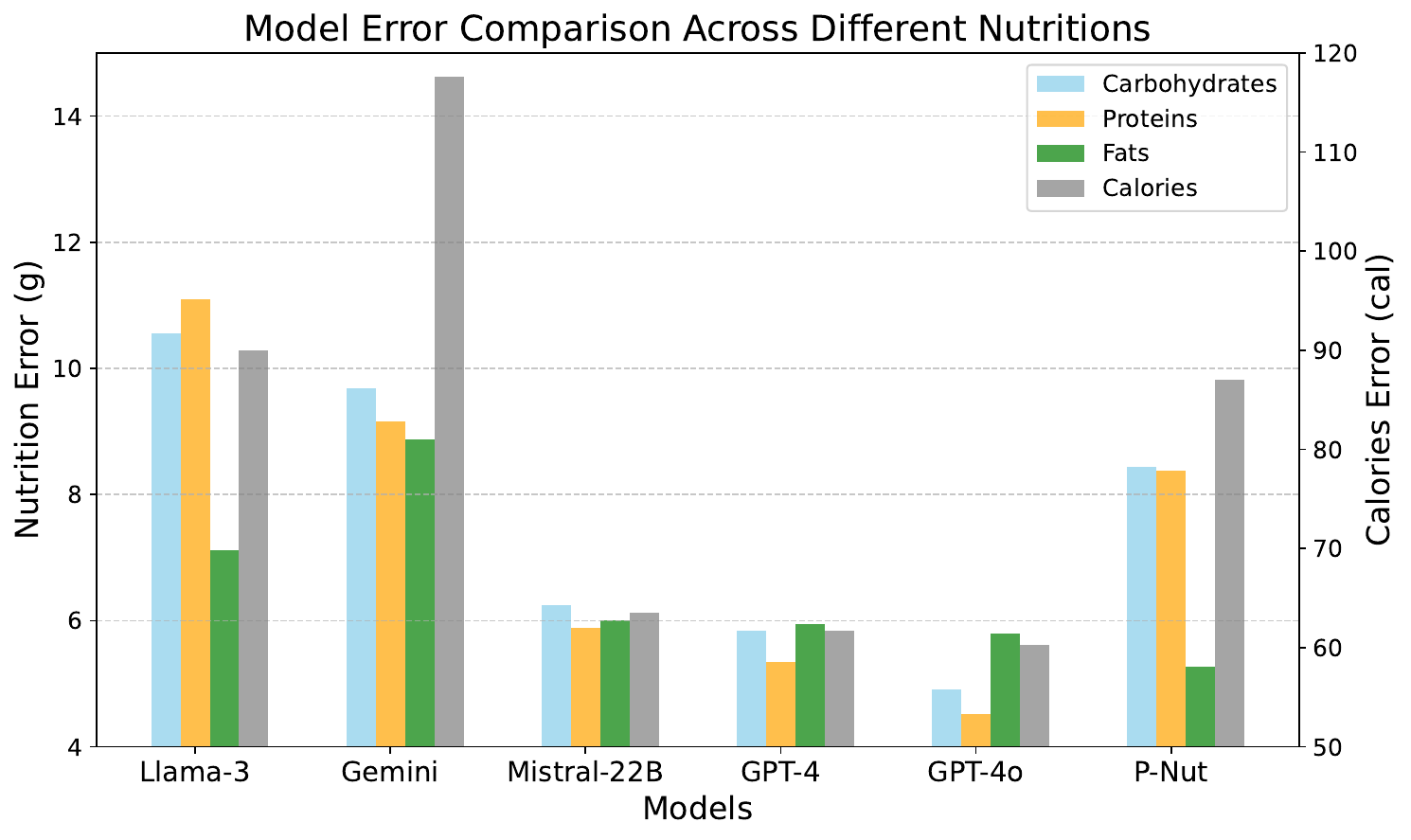}
    \caption{Performance of different models for the dietary analysis.}
    \label{error_llm}
\end{figure}

\subsection{Dietary Analysis}
We use the record of home-cooked dishes in the BOOHEE dataset to evaluate the capability of tailored LLMs for the dietary analysis module.
We compare the performance of several popular LLMs, including GPT-4 \cite{achiam2023gpt}, 
GPT-4o \cite{gpt4o2024}, Mistral-22B \cite{jiang2024mixtral}, Llama3 \cite{touvron2023llama}, Gemini \cite{team2023gemini} and one existing work P-Nut \cite{ispirova2020p}. ChatGPT cannot analyze the dietary in the experiment and thus is excluded from the experiment.
In the experiment, tailored LLMs are utilized to analyze the content of carbohydrates, calories, proteins, and fats in the patient's dietary intake. We use 30\% of the BOOHEE dataset to train P-Nut, and the rest 70\% is used for test.

Figure \ref{error_llm} compares the MAE (in gram) of the six models in estimating the content of the four nutrients in meals. Among them, Mistral-22B, GPT-4 and GPT-4o achieve similar performance with lower estimation error comparing to the other two models. These three models are all good at estimating the content of calories. On the other hand, Gemini performs worst in estimating calories while Llama3 and P-Nut are slightly better than Gemini in general but not much. We see although P-Nut is particularly trained on the dataset, it still lacks the extensive knowledge base to provide accurate analysis.
\begin{figure}
    \centering
    \includegraphics[width=0.44\textwidth]{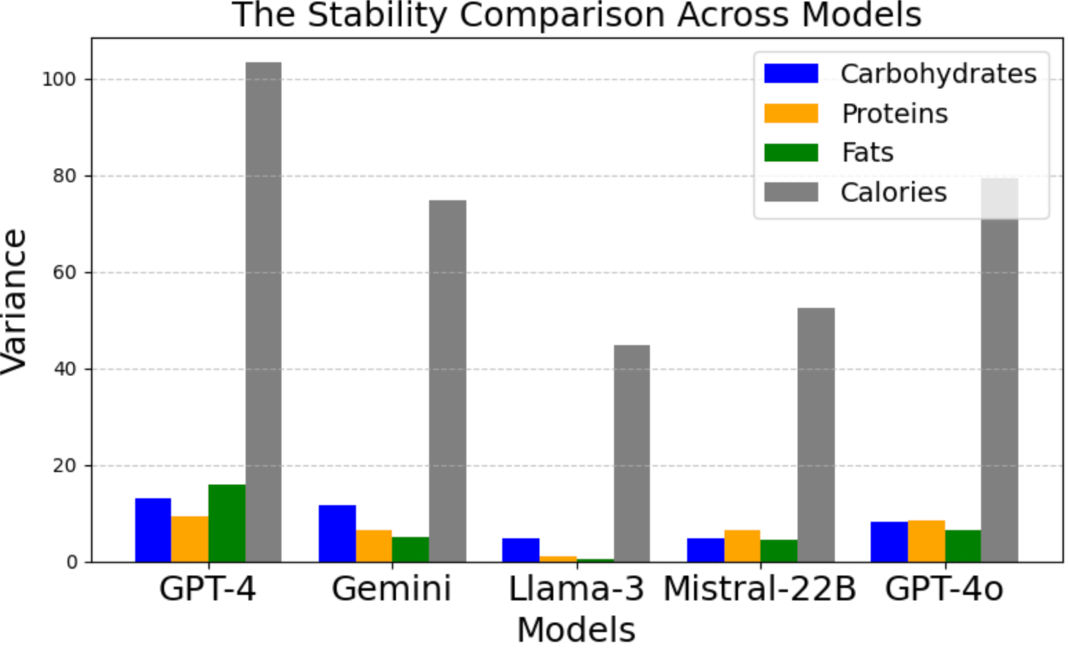}
    \caption{The performance stability of different LLMs.}
    \label{stability}
\end{figure}
Figure \ref{stability} reports the error variance of the nutrient estimation results of LLMs. Due to the variability where the same inputs to LLMs can generate different outputs, stability has emerged as a crucial factor to consider.
We see Llama3 achieves lower variance of the estimation error, and Mistral-22B is slightly worse but not much. The other three models, i.e., GPT-4, GPT-4o and Gemini, exhibit similar performance in terms of error variance.
In our implementation, we choose GPT-4o due to its highest estimation accuracy on nutrition estimation, which is one of the most important factors for glucose and insulin prediction.

\subsection{Performance of Insulin Delivery Strategy Determination}\label{titra_com}
\begin{figure}
    \centering
    \includegraphics[width=0.44\textwidth]{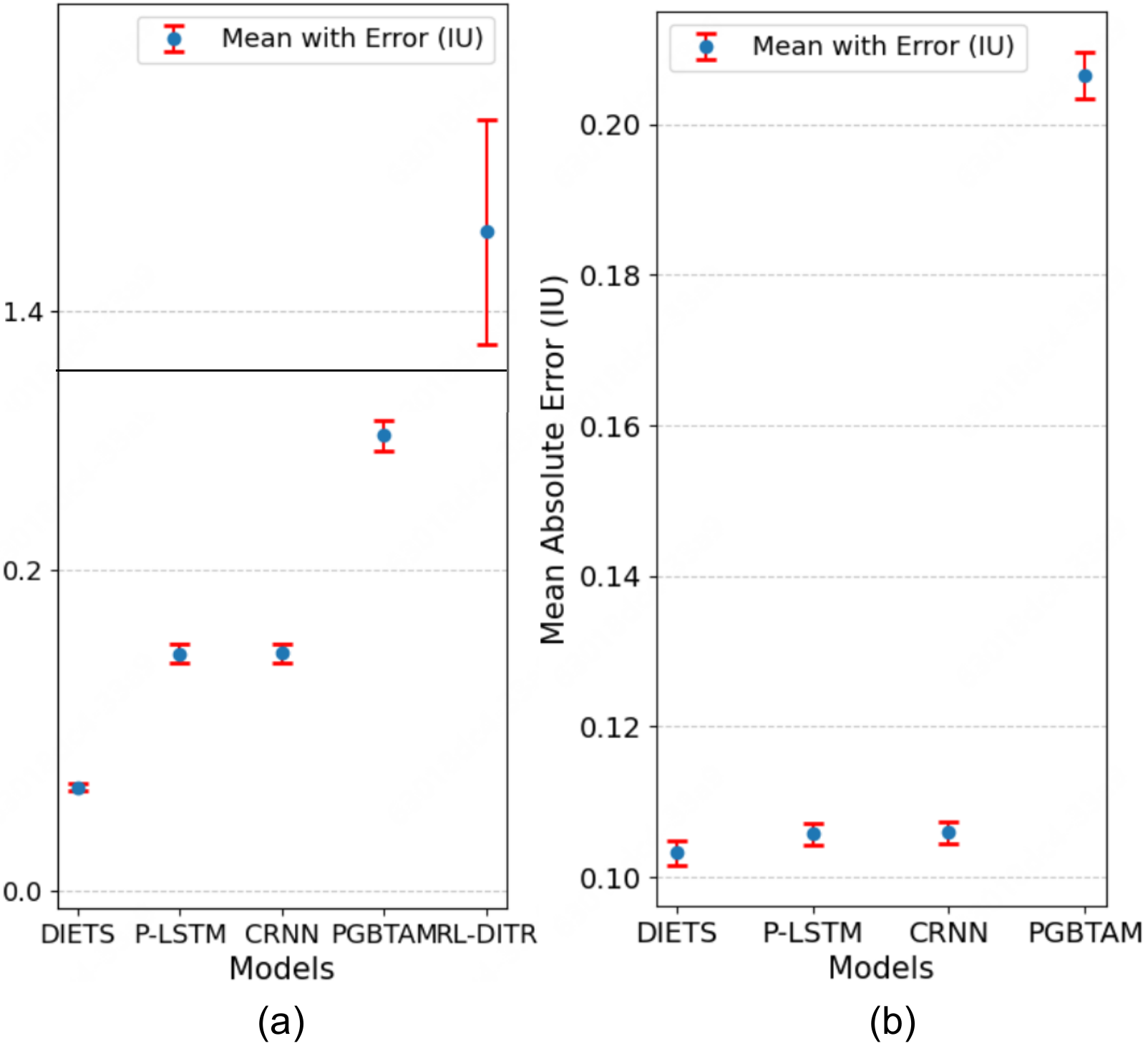}
    \caption{Performance of insulin delivery strategy determination on (a) ShanghaiDM dataset and (b) OhioT1DM dataset}
    \label{error_bar}
\end{figure}

We compare the performance of the titration model in DIETS with related existing models discussed in Section \ref{subsect:experiment_details} on ShanghaiDM Dataset.
The experiment setting is as follows. In the test dataset, for the given expected blood glucose levels, we examine whether the models can determine similar glucose delivery strategies with the ground-truth recorded in the dataset.
In addition to the RL-DITR, we also adapt the glucose prediction models, PGBTAM, CRNN and P-LSTM for the titration task. Some minor modifications were made to their data inputs and outputs, while preserving the original structure to maintain their capability to learn the effects of insulin on the endocrine system. These adaptations allow the models to shift their focus from predicting future blood glucose levels affected by insulin injections to calculating insulin dosages responsible for specific glucose variations. 
In other words, the model calculates the insulin dosing required to produce the observed fluctuations in blood glucose levels.
As all of these existing works assume having prior knowledge about the nutrients in meals, we use the output of our LLM-based dietary analysis module as the input for all models investigated in the experiments for comparison.
The shared inputs among all the models include historical blood glucose levels, carbohydrate intake, bolus, and basal insulin injection, and future expected blood glucose levels. 
All the models are trained with the same training set and tested on the same test set. The result is reported in Figure \ref{error_bar}(a). We tested the RL-DITR in \cite{wang2023optimized} directly in the test set, for the professional instructions from clinicians are not accessible in ShanghaiDM dataset. 

To validate the generalizability of DIETS, experiments were also conducted using the ohioT1DM dataset. Unlike the Shanghai Diabetic dataset, the ohioT1DM dataset lacks personalized patient information, creating distinct conditions for each dataset. As discussed in section 3, to achieve favorable experimental outcomes in scenarios devoid of personal information, the DIETS-LSTM model was developed as an adaptation. And the results are shown in Figure \ref{error_bar}(b). As RL-DITR requires the patient's personal details as input, such as age, gender and BMI, which are not provided by the ohioT1DM dataset, it is not included in this experiment.


As the results suggest, on the Shanghai Diabetics Dataset, the DIETS achieves the best performance on determining insulin delivery strategies to reach the expected blood glucose levels. The MAE of the estimated insulin injection is 0.0641 $IU$, which is at least 50\% less than the error from all other models, and DIETS is also the most stable model among all competitors.
The RL-based model RL-DITR performs the worst. This is probably due to the poor adaptation capacity of the model. 
In the ohioT1DM dataset scenario, DIETS remained the best-performing and most stable model. However, due to the lack of personalized patient information available for customized analysis, DIETS's lead was not as pronounced compared to previous datasets.

\subsection{Performance of Glucose Prediction}
\begin{figure}
    \centering
    \includegraphics[width=0.42\textwidth]{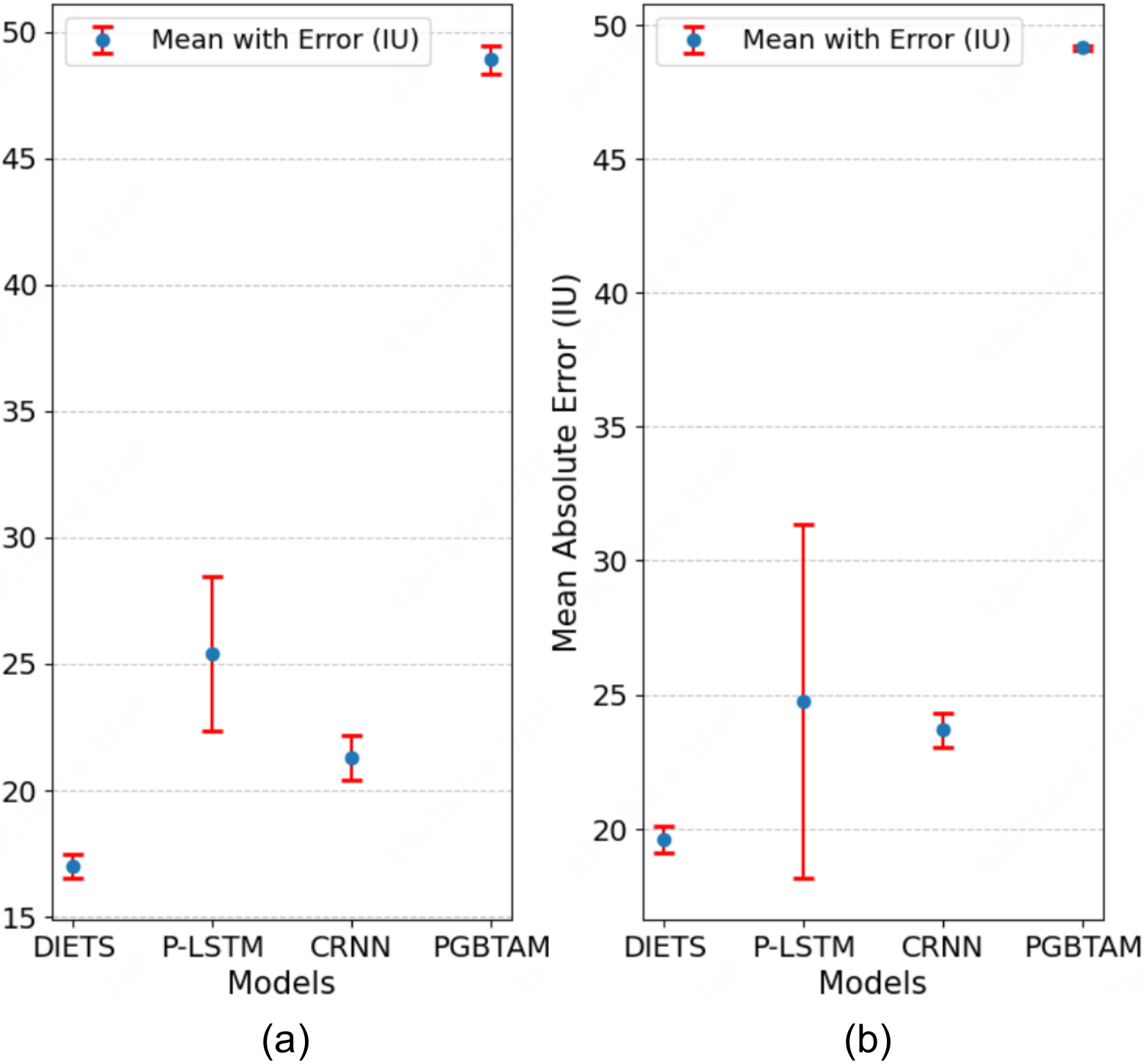}
    \caption{Performance of glucose prediction on (a) ShanghaiDM dataset and (b) OhioT1DM dataset.}
    \label{error_bar_glycemic}
\end{figure}

We compare the performance of the glucose forecasting model of DIETS with three state-of-the-art models mentioned in Section \ref{subsect:experiment_details}, i.e., PGBTAM, CRNN and P-LSTM. 
All models are originally designed for blood glucose prediction. 
The experiment setting is as follows.
We also evaluate models on both ShangHai Diabetic dataset and ohioT1DM Dataset.
All models are used to predict the blood glucose variations in the next 2 hours. The inputs to all models are the same, including the historical blood glucose variation, patient's basic information, and the planned insulin delivery strategies, etc.
The metric is the MAE of the predicted blood glucose value ($mg/dl$), and the results are reported in Figure \ref{error_bar_glycemic}.
As the results suggest, the glucose prediction model in DIETS achieves the best performance in glucose prediction with an average MAE of 15.91 $mg/dl$ in Shanghai Diabetic Dataset, and 19.60 $mg/dl$ in ohioT1DM Dataset, significantly outperforms SOTA models. And the DIETS is comparatively the most stable model among all competitors.

\begin{table*}[h!]
\centering
\caption{Evaluation of titration performance across various fine-tuning options and training data volumes (MAE)}.
\label{foundation}
\begin{tabular}{c|c|c|c|c|c}
\hline
\textbf{Amount of training data} & \textbf{Single} & \textbf{Foundation} & \textbf{FT-full} & \textbf{FT-CNN\&Dense} & \textbf{FT-Dense} \\ \hline
1 day   & 0.1324 & \multirow{4}{*}{0.0641} & 0.1283  & 0.1258   & \textbf{0.1254}   \\ \cline{1-2} \cline{4-6}
3 days  & 0.0758 &                          & 0.0923  & 0.0677   & \textbf{0.0598} \\ \cline{1-2} \cline{4-6}
6 days  & 0.1016 &                          & \textbf{0.0635} & 0.1180   & 0.0705   \\ \cline{1-2} \cline{4-6}
10 days & 0.0780 &                          & 0.0764  & 0.0884   & \textbf{0.0551} \\ \hline
\end{tabular}
\end{table*}

\subsection{Ablation Study on the Incorporated Features}
\label{subsect:ablation_study}

In our study, both the insulin injection strategy determination model and the glucose prediction model incorporate numerous features. Evaluating the efficiency of these features and their impact on model accuracy remains a critical consideration.
We conduct a series of experiments to investigate their impact on the effectiveness of the two models in Shanghai Diabetic Dataset. We train the two models using different groups of features and test their performances.
This approach allows us to systematically evaluate and refine the feature selection process, enhancing the overall performance and reliability of our design.
Nine groups of features are investigated in the experiments. Their configurations are as follows.

\begin{figure}
    \centering
    \includegraphics[width=0.42\textwidth]{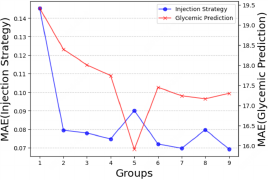}
    \caption{The impact of different features on the titraion model and glucose prediction model.}
    \label{ablation}
\end{figure}

\noindent$\bullet$\hspace{0.5em}  \textbf{G1}: Blood glucose levels, insulin injection dosages. \\
\noindent$\bullet$\hspace{0.5em}  \textbf{G2}: Features in G1, intake of carbohydrates.\\
\noindent$\bullet$\hspace{0.5em}  \textbf{G3}: Features in G2, intake of calories, proteins, and fats. \\
\noindent$\bullet$\hspace{0.5em}  \textbf{G4}: Features in G3, intake of anti-glycemic drugs.\\
\noindent$\bullet$\hspace{0.5em}  \textbf{G5}: Features in G4, basic information of patients.\\
\noindent$\bullet$\hspace{0.5em}  \textbf{G6}: Features in G4, historical 24-hour basal injection.\\
\noindent$\bullet$\hspace{0.5em}  \textbf{G7}: Features in G4, basic information of patients, historical 24-hour basal injection.\\
\noindent$\bullet$\hspace{0.5em}  \textbf{G8}: Features in G6, medical record.\\
\noindent$\bullet$\hspace{0.5em}  \textbf{G9}: Features in G7, medical record.


The results are plotted in Figure \ref{ablation}.
We see the features have different impact on the two models. 
For the titration model, the feature group G7 and G9 achieve a comparable performance, which is better than other groups. Although the model gets slightly improved with the addition of medical records of patients in G9, accessing medical records typically requires hospital visits and medical examinations, posing a significant barrier to data collection and complicating the use of the DIETS framework for patients with limited access to healthcare. 
For the glucose prediction model, the feature group G5 achieves the best performance, which indicates the effectiveness of the individual's basic information.
Consequently, we utilize features from G7 as the model inputs for the insulin titration model, and features in G5 for the glucose prediction model.

\subsection{Impact of Personalized Fine-tuning}
In DIETS, we employ two-stage training process to develop personalized titration model for each individual. In this section, we investigate the impact of different fine-tuning alternatives on the final model performance. 
We compared the performance of the following models.

\noindent$\bullet$\hspace{0.5em}  \textbf{Single}: The titration model is only trained by the data of the individual, without using public datasets.\\
\noindent$\bullet$\hspace{0.5em}  \textbf{Foundation}: The foundation model trained in the first stage. No fine-tuning is applied.\\
\noindent$\bullet$\hspace{0.5em}  \textbf{FT-full}: The titration model is fine-tuned from the foundation model. During the fine-tuning process, all the parameters in the DIETS are trainable.\\
\noindent$\bullet$\hspace{0.5em}  \textbf{FT-CNN\&Dense}: The titration model is fine-tuned from the foundation model. During the fine-tuning process, only the parameters in the Convid 1d layer and dense layer are trainable.\\
\noindent$\bullet$\hspace{0.5em}  \textbf{FT-Dense}: The titration model is fine-tuned from the foundation model. During the fine-tuning process, only the parameters in the dense layer are trainable.

In this experiment, we also investigate the impact of different amounts of training data. We train the above models using three subsets of public datasets, i.e., 1-day, 3-day, 6-day, and 10-day dataset, and the results are summarized in Table \ref{foundation}.
The results reveal that model performance does not strongly correlate with the volume of training data; notably, three days of data are sufficient for training DIETS. 
The Single model without public datasets cannot perform well with all four training datasets.
FT-Dense model performs the best with 1-day, 3-day and 10-day training data. FT-full model performs best with the 6-day training data, which is slightly better than FT-Dense model. The result suggest that completely fine-tuning the model in the second-stage would hinder the model performance. 
In comparison, only fine-tuning the dense layer of the foundation model achieves the best performance. In comparison with the foundation model, it advised that personalized models be trained using at least three days of a patient's glycemic data. Or the foundation model should be utilized directly to provide injection strategy recommendations.

\begin{table}[!]
\centering
\caption{Performance on Type 1 and Type 2 Diabetes.}
\label{mae_type_c}
\begin{tabular}{|c|c|c|c|c|}
\hline
\textbf{Trained by } & \textbf{Type 1} & \textbf{Type 2} & \multicolumn{2}{c|}{\textbf{Type 1 \& 2}} \\ \hline
\textbf{Tested on}& \textbf{Type 1}&\textbf{Type 2} & \textbf{Type 1} & \textbf{Type 2} \\ \hline
\textbf{MAE ($IU$)}&0.0731 & 0.0745 & 0.0635 & 0.0650 \\ \hline
\end{tabular}
\end{table}

\subsection{Performance on Type 1 and Type 2 Diabetes}
We compare the performance of the titration models on different types of diabetes. We make use of two sub-datasets, i.e., Type 1 dataset with data records from 11 people with type 1 diabetes, and Type 2 dataset with data records from 20 people with type 2 diabetes.
We train the model using three settings respectively, i.e., the Type 1 dataset, the Type 2 dataset, and both datasets. We compare the three models and report the results in Table \ref{mae_type_c}.
The test results show that when the model is trained exclusively on either Type 1 or Type 2 diabetes datasets, its performance is better for Type 1 patients compared to Type 2. This may be attributed to the greater complexity and individual variability of Type 2 diabetes. However, when the model is trained on a combined dataset of both Type 1 and Type 2 diabetes, there is an improvement in performance across both datasets. This indicates that despite their differences, there are commonalities between the two types of diabetes from which the model can learn useful insights from each other.

\subsection{Impact of Sliding Window Size }
As mentioned in Section \ref{dp}, the input data is partitioned by a sliding window, and every data clip consists of future duration and previous duration. The future duration of the input data is fixed to be 2 hours (8 time slots), while the previous duration of input data is still flexible. For example, if the sliding window size is 6 hours, then previous duration of data is 4 hours. 
To understand the impact, we compare the performance of titration models with different sliding window sizes for data partitioning, including 8 hours (32 time slots), 6 hours (24 time slots), and 4 hours (16 time slots).
The performance is reported in Table \ref{mae_time}. It is evident that the model with the biggest window size, i.e., longest input gets the lowest error. Thus in DIETS, we use the window size of 8 hours to train the titraion model and the glucose prediction model. 

\begin{table}[!]
\centering
\caption{Impact of the the sliding window size.}
\label{mae_time}
\begin{tabular}{|c|c|c|c|}
\hline
total Duration & 8 hours &6 hours & 4 hours\\ \hline
MAE($IU$) & 0.0641 & 0.0827 & 0.0724 \\ \hline
\end{tabular}
\end{table}

\subsection{Model Size and Overhead}
The titration model, with an allocated memory of 201.4 MB while inference, contains 11,233,692 parameters. The average inference time per
instance, based on 10000 inferences, is 25.4ms. The glucose prediction model contains 11,182,384 parameters, with an allocated memory of 113.1 MB. The average inference time per instance, based on 10000 inferences, is 17ms.
As for the dietary analysis module, which relys on LLM cloud servers, it does not have any memory occupation. The average time consumption for this module is 0.85s, based on 100 inferences.
Most modern smartphones like the Google Pixel 7 (Processor: Google Tensor G2, RAM: 8GB) are equipped with adequate processors and sufficient memory to effectively handle the whole diabetes management system inference.

\section{Related Work}
\label{sect:related_work}

\textbf{Nutrient analysis}.
Most existing research on blood glucose management either assume having prior knowledge about the nutrients of the patient's meals \cite{brown2018first, silva2022real} or require the patient to adhere to a predefined set of food intake \cite{emerson2023offline,jaloli2024basal}.
Some research works \cite{ispirova2020p,ispirova2024msgen} use unsupervised learning and supervised learning NLP models to estimate nutrients. However, it is hard for those traditional NLP models to handle the unstructured and casual textual descriptions from the patients. Due to the diversity of patients and the corresponding variety in their diets, traditional NLP models often lack the extensive knowledge base to provide accurate analysis.

\textbf{Blood glucose management}.
There are two categories of works for glucose management: Traditional control theory-based methods and machine learning approaches. 
MPC-based systems use mathematical models based on inputs like glucose concentration and insulin infusion rates to adjust parameters for optimal dosing \cite{nwokolo2023artificial}.
PID-based systems, on the other hand, calculate insulin doses by assessing deviations from target glucose levels, the actual difference between current and target glucose, and the rate of glucose change \cite{nwokolo2023artificial}. 
Fuzzy logic systems are employed to replicate the decision-making processes typical of diabetes clinicians \cite{nwokolo2023artificial}. These methods, however, are generally limited to minor adjustments in basal insulin, and suffer from the significant inherent delay in insulin’s impact on blood glucose levels. 
Machine learning models, primarily the reinforcement learning (RL), have been studied for insulin titration and glucose prediction. 
Jaloli \textit{et al.}  \cite{jaloli2024basal} and Emerson \textit{et al.}  \cite{emerson2023offline} exemplify using simulators to develop RL-based systems without endangering human subjects. \cite{jaloli2024basal} utilized a multi-agent RL model for constant trial-and-error until the model learned good knowledge in insulin titration for both basal and bolus. In \cite{emerson2023offline}, an offline RL model is employed to train on the medical datasets and simulator. 
\cite{wang2023optimized} proposed a glycemic control system with the RL model where professional clinicians evaluate the feedback of the decision-making of the RL model.
However, the training of RL-based systems highly relies on the availability of the extensive feedback from professionals, which is impractical for everyday life, and they struggle with adapting the model to new patients due to individual differences among patients.

\textbf{Blood glucose prediction}.
In practice, glucose prediction is an important component for diabetes management to prevent potential risks. Existing works primarily make use of ineffective machine learning models. 
Aliberti \textit{et al.} \cite{aliberti2019multi} use LSTM networks for blood glucose prediction.
Li \textit{et al.} \cite{li2019convolutional} utilize the convolutional neural network to forecast glucose levels for simulated patient cases.
Yang \textit{et al.} \cite{yang2023short} propose a short-term prediction method of blood glucose based on temporal multi-head attention mechanism for diabetic patients. 
Cheng \textit{et al.} \cite{Cheng2024TowardSG} propose an end-to-end pipeline for short-term glucose prediction solely based on CGM time series data.
Cai \textit{et al.} \cite{cai2021bayesian} use the Gaussian process regression in blood glucose prediction.
All of those models lack the capacity of effectively predicting accurate blood glucose levels due to the lack of personalized fine-tuning.

\section{Conclusion}

We introduce DIETS, the first comprehensive insulin management framework for people with diabetes in everyday life, without the need for expert supervision. Effective insulin delivery recommendations based on dietary information analyzed by tailored LLM are developed by DIETS to control future blood glucose levels. Additionally, DIETS also incorporates a glucose prediction model to prevent potential risks of hyperglycemia or hypoglycemia. Extensive experiments on various public datasets demonstrate the effectiveness of DIETS, indicating its high potential for clinical trials.

\bibliographystyle{ACM-Reference-Format}
\bibliography{sample}

\appendix

\end{document}